# On idealism of Anton Zeilinger's information interpretation of quantum mechanics

Francois-Igor Pris (И. Е. Прись)[1]

*Abstract*

We argue that Anton Zeilinger's «foundational conceptual principle» for quantum mechanics according to which an elementary system carries one bit of information is an idealistic principle, which should be replaced by a realistic principle of contextuality. Specific properties of quantum systems are a consequence of impossibility to speak about them without reference to the tools of their observation/identification and, consequently, context in which these tools are applied.

## 1. Introduction

In contemporary physics and, in particular, within some information interpretations of quantum mechanics, the idea that information is a fundamental characteristic of the universe has become widespread. It is also said that the universe (nature) is a huge computer or program. Similarly, in the footsteps of Galileo, modern scientists believed that the book of nature is written in the language of mathematics.

From the perspective of working science, there is nothing wrong with this kind of projection of models onto reality, since the fundamental norm of science, such as physics, is truth. But from a philosophical point of view, conflating the ideal (conceptual) – the models are ideal – and the real (things themselves) is unacceptable. This is one of the reasons why philosophical discussions arise. Discussions regarding the interpretations of quantum mechanics, as we know, have been going on for almost a hundred years.

In this paper, from the point of view of a Wittgensteinian contextual realism which distinguishes between the categories of the real and the ideal, we analyze Anton Zeilinger's «information principle» which he considers as a «foundational conceptual principle» for quantum mechanics (Zeilinger 1999). We argue that this is an idealistic principle, which should be replaced by a realistic principle of contextuality.

## 2. Contextual realism

From the perspective of a contextual realism, as we understand it, concepts (rules, norms, thought, truth, knowledge, information) belong to the category of the ideal (Benoist 2017). If they were real, they could not grasp reality, and it would need something «unreal» to grasp them themselves as a special reality. They are «real» only in the sense that they are rooted in reality, have real conditions for their existence and their effectiveness (application). In the process of cognition, we apply concepts, make judgments (always in a context). The ideal cannot affect the knowable. Thus, it is necessary to reject subjectivistic interpretations of quantum mechanics, according to which the observer influences the observed quantum system. It is also wrong to speak of an insurmountable correlation between the quantum observer and the observed quantum system, as if it were not the quantum nature itself but only the quantum things-for-us that would be available to us.

According to our Wittgensteinian contextual realism, every established and confirmed scientific theory, including quantum mechanics, has a domain of its applicability. Within this domain it plays the role of a Wittgensteinian rule/norm for measuring reality, has a logical certainty. As a consequence, it is not falsifiable. The structure of the quantum measurement problem is that of the Wittgensteinian rule-following problem. The gap between quantum formalism and reality is closed within a language game of its application, that is in practice. The so-called reduction of the wave function is not a physical process. In a sense, our view is a return to the Copenhagen interpretation of quantum mechanics, corrected within the framework of contextual realism (Pris 2020).[2]

[1] frigpr@gmail.com

### 3. The information principle

Anton Zeilinger proposed an «information principle» as a «foundational conceptual principle» for quantum mechanics [1]. From the point of view of the contextual realism, we argue that this is an idealistic principle, which should be replaced by a realistic principle of contextuality.

The Austrian physicist proposes the following principle: (1) «An elementary system carries 1 bit of information» (Zeilinger 1999, 635). However, he notes that this formulation can be regarded as a *definition* of an elementary physical system. Equivalently – and this is important for understanding our position – the fundamental principle can be reformulated in linguistic terms: (2) «An elementary system represents the truth value of one proposition» (Zeilinger 1999, 635). We would reformulate (2) as follows: (2') «An elementary system is described/represented by one (true) proposition». The connection between (1), (2) and (2') is obvious: 1 bit of information is expressed as one true proposition; therefore, the system with which a true proposition is associated is a system that «carries» 1 bit of information.

Based on the information principle, Zeilinger derives properties of quantum systems. We accept his theoretical derivation. At the same time, we believe that the philosophical position of the scientist contains contradictions and conceptual confusions.

For example, Zeilinger makes the following distinction between the classical and quantum approaches: «Therefore, while in a classical worldview, reality is a primary concept prior to and independent of observation with all its properties, in the emerging view of quantum mechanics the notions of reality and of information are on an equal footing» (Zeilinger 1999, 642).

The premise is wrong. The concept of reality is unambiguous; it is the same for any field of science: «reality is *just what it is*» (Benoist 2014, 220. By definition it cannot be «on an equal footing» with information, which, if understood as knowledge, belongs to the category of the ideal (information is knowledge «about reality», not a (special) part of reality). Contradicting himself, Zeilinger writes: «We have knowledge, i.e., information, of an object only through observation» (Zeilinger 1999, 633).

The «classical worldview» is quite correct in considering «reality» as a primary concept. Another issue is that traditional metaphysical realism, which asserts the «givenness» of the external reality with determinate properties – the internal properties of autonomous (decontextualized, absolutized) «external» objects – is a wrong position. One can speak of a such kind of intrinsically meaningful reality only secondarily, as a domain of reality which is already known, «domesticated», conceptualized. The things which are already known (identified), carry, so to speak, information about themselves. In fact, one can agree with A. Zeilinger when he writes that «in physics we cannot talk about reality independent of what can be said about reality», but only if this formulation is understood not in the sense of linguistic idealism (language is primary, reality is secondary and corresponds to (or is constructed by) language) or correlationism (language in itself and reality in itself have no meaning or are at any rate unknowable; there are only insurmountable interpenetrable «correlations» between language and reality), but as a tautology: we cannot really talk (think) about reality (real things) unless we can talk (think) about it (them), that is, unless we apply suitable means to identify it (them). This is the logical (therapeutic) view of contextual realism. Language «correlates» with reality only in the sense that it is able to describe reality (the things themselves) adequately in context; it allows us to make sense of it. It is in this way – as a critique of metaphysical realism and correlationism, rather than in the neo-Kantian sense – that we also propose to interpret

---

[2] Our view that quantum theory is a Wittgensteinian rule/norm has some similarity with QBism. For QBism, Born's rule and quantum theory have the normative character; they do not describe the objective (external) reality, existing independently from the subject and the language use – the ordinary as well as the theoretical (Fuchs 2019). However, QBism considers the quantum measurement as an interaction of the subject with a quantum system, allowing to the former to affect the latter, reality. So, for QBism, only the result of an interaction is measured and not the real things as they actually exist independently from the subject. This is rather an anti-realist, not a realist approach.

Niels Bohr's words quoted by Zeilinger: «It is wrong to think that the task of physics is to find out how Nature is. Physics concerns what we can say about Nature» (quoted in (Zeilinger 1999, 637)). In fact, it is our contextually correct statements about Nature that show how Nature is.

## 4. From the information principle to the principle of contextuality

Zeilinger's informational derivation assumes an interpenetration of information and reality (here, it seems to us, there is a direct parallel with Hegel's idealism), as well as conservation of information. As a consequence, the absence of information amounts to the absence of reality: hence quantization of physical quantities, irreducibility of quantum randomness, impossibility of introduction of hidden variables, «non-locality» of quantum correlation, principle of complementarity, and so on (true consequences of a false assumption). The problem is that quantum systems, carrying information about themselves, turn out to be autonomous (absolutized) systems.

Instead we propose to consider quantum systems and their characteristics as identifiable in context. Quantum ontology (in fact any ontology) and cognition/knowledge are contextual (Pris 2020).

Identification (knowing) in context is expressed in the form of a propositional knowledge. The elementary expression is a proposition. Hence, it follows from the principle of contextuality, that an elementary physical system can be identified/represented in the form of a proposition (1 bit of information). For Zeilinger, on the contrary, a physical system represents propositional content and (carries in itself) information. At the same time, contradicting himself, he explains that «notions such as that a system "represents" the truth value of a proposition or that it "carries" one bit of information only implies a statement concerning what can be said about possible measurement results» (Zeilinger 1999, 635).

The strange properties of quantum systems are a consequence of impossibility to speak about them without reference to the means of their observation/identification and, consequently, context in which these means are applied (in general, according to the Destouches-Février theorem, theories, within which phenomena are not separable from a way of access to them, are essentially probabilistic (Destouches-Février 1951)). In particular, to explain the quantum correlation the assumption of its nonlocality is not required. Correlating quantum events are related to each other in a causal way. But it is not classical, but quantum causality, expressed by an entangled wave function. This or that particular correlation does not arise in a measurement; in a measurement, it is identified in the measurement context. In contrast to Zeilinger's principle of quantization of information, the principle of contextuality explains it realistically.

## 5. Conclusion

Thus, in our view, Zeilinger's philosophical position can and should be turned right-side up. As a result, it turns out that the foundational conceptual principle of quantum physics is not the principle of information, but the principle of contextuality. Only within the framework of contextual point of view Zeilinger's informational derivation of quantum properties and phenomena gets a truly realistic interpretation.

# Об идеализме информационной интерпретации квантовой механики Антона Цайлингера

*Аннотация*

Мы утверждаем, что предложенный А. Цайлингером «фундаментальный концептуальный принцип» квантовой механики, согласно которому элементарная система является носителем одного бита информации, является идеалистическим принципом, который должен быть заменён реалистическим принципом контекстуальности. Специфические свойства квантовых систем – следствие невозможности говорить о них безотносительно к средствам их идентификации и, следовательно, контексту, в котором эти средства применяются. В частности, для объяснения квантовой корреляции не требуется предположение о её нелокальности. Коррелирующие квантовые события связаны между собой причинным образом. Но это не классическая, а квантовая причинность, выражаемая запутанной волновой функцией. Та или иная конкретная корреляция не возникает при измерении; при измерении она идентифицируется в контексте. В отличие от предложенного Цайлингером принципа квантования информации, принцип контекстуальности объясняет её реалистически.

*Ключевые слова*: квантовая механика, информация, идеализм, реализм, контекст, следование правилу, квантовая корреляция